\documentclass[pra,twocolumn,longbibliography,superscriptaddress,10pt]{revtex4-1}
\usepackage{graphicx}
\usepackage{dcolumn}
\usepackage{color}
\usepackage{times}
\usepackage{bm}
\usepackage{amssymb}
\usepackage{amsmath}
\usepackage{epsfig}
\usepackage{epstopdf}
\usepackage{dsfont}
\usepackage{subfigure}
\usepackage{tikz}
\usepackage{CJK}
\usepackage{indentfirst}
\usepackage{cases}
\usepackage[colorlinks, citecolor=blue, linkcolor=blue, urlcolor=blue]{hyperref}
\usepackage[mathscr]{euscript}
\usepackage{enumerate}
\usepackage{amsthm}
\usepackage{overpic}

\makeatletter
\renewcommand{\maketag@@@}[1]{\hbox{\m@th\normalsize\normalfont#1}}
\makeatother

\begin{document}
\renewcommand{\thefootnote}{\fnsymbol {footnote}}

\title{Entropic uncertainty relations  in  {Schwarzschild} space-time}

\author{Tian-Yu Wang}
\affiliation{School of Physics \& Optoelectronic Engineering, Anhui University, Hefei 230601,  People's Republic of China}

\author{Dong Wang} \email{dwang@ahu.edu.cn}
\affiliation{School of Physics \& Optoelectronic Engineering, Anhui University, Hefei 230601,  People's Republic of China}


\begin{abstract}{
The uncertainty principle is deemed as one of cornerstones in quantum mechanics, and exploring its lower limit of uncertainty will be helpful to understand the principle's nature.
In this study, we propose a generalized entropic uncertainty relation for arbitrary multiple-observable in multipartite system, and  further derive a  tighter lower bound by considering Holevo quality and mutual information. Importantly, we detailedly discuss the proposed uncertainty relations and quantum coherence in the context of {Schwarzschild} space-time. It is interesting to find  that Hawking radiation will damage the coherence of the physically accessible region and increase the uncertainty. Furthermore, we argue that the properties of the uncertainty in {Schwarzschild} space-time can be explained from  the systems' purity and the information redistribution  of the different regions.
Therefore, it is believed that our findings  provide the generalized entropic uncertainty relations in multipartite systems, which may facilitate us deeper understanding of   quantumness and information paradox of the black holes.}
\end{abstract}

\maketitle
\section{Introduction}

The uncertainty relation is at the heart of quantum mechanics, which demonstrates that one is unable to obtain the exact values of two non-commutative observables at the same time, other than those in classical counterpart. The concept of uncertainty relation was first proposed by Heisenberg \cite{WH}, and then Kennard \cite{EHK} strictly proved that the standard deviation can be used to depict the uncertainty relation between momentum and position:
\begin{equation}
\begin{split}
 \Delta x \Delta p \ge \frac{\hbar }{2},
\label{Eq.1.1}
\end{split}
\end{equation}
where $\Delta  x$ ($\Delta  p$) represents the standard deviation of position (momentum). Based on this theory, Robertson \cite{HPR} generalizes it to the relation for arbitrary two non-commutating observables.

Subsequently, some scholars have found that entropy is more suitable for describing the uncertainty relation, because it can get rid of some restriction from the standard-deviation-based uncertainty relation. Everett \cite{HE} and Hirschman \cite{IIH} originally brought information entropy into the uncertainty principle in 1957. Later on, Deutsch \cite{DD} proposed the celebrated entropic uncertainty relation (EUR) via Shannon entropy, i.e.,
\begin{equation}
\begin{split}
H(\hat{R})+H(\hat{Q})\ge \log_2 \left ( \frac{2}{1+\sqrt{c\left ( \hat{R},\hat{Q} \right ) } }  \right ) ^2,
\label{Eq.1.2}
\end{split}
\end{equation}
where $H(\hat{X})=-\sum_i p_i \log _2 p_i$ stands for Shannon entropy with $p_i$ representing the probability of the $i$-th outcome. $c\left ( \hat{R},\hat{Q} \right )= \underset{i,j}{\max} \left | \left \langle {\psi_i^{\hat{R}} } \middle| {\psi_j^{\hat{Q}}}  \right \rangle  \right |^2 $ is the maximum overlap between $\hat{R}$ and $\hat{Q}$ with $\langle {\psi_i^{\hat{R}} }|$ ($ | {\psi_j^{\hat{Q}}} \rangle $) represents the eigenvector of observable $\hat{R}$ ($\hat{Q}$).
Based on this, Kraus \cite{KK} and Maassen and Uffink \cite{HM} proposed a more simplified EUR:
\begin{equation}
\begin{split}
H(\hat{R})+H(\hat{Q})\ge - \log_2 c \left ( \hat{R},\hat{Q} \right ) =: q_{MU}.
\label{Eq.1.3}
\end{split}
\end{equation}

Furthermore, some researchers have shown that the uncertainty can be reduced when the measured subsystem is correlated with another \cite{JMR,MB}. To be specific, the novel EURs with quantum memories was presented by Renes and  Boileau \cite{JMR}, and Berta \emph {et al.} \cite{MB}:
\begin{equation}
\begin{split}
S( \hat{R} | B ) + S( \hat{Q} | B ) \geqslant q_{MU} + S(A|B),
\label{Eq.1.4}
\end{split}
\end{equation}
and
\begin{equation}
\begin{split}
S( \hat{R} | B ) + S( \hat{Q} | C ) \geqslant q_{MU},
\label{Eq.1.5}
\end{split}
\end{equation}
where $S( \hat{R} | B ) = S(\rho_{\hat{R}B}) - S(\rho_{B})$  is the conditional von Neumann entropy of the system after $\hat{R}$  measurement of particle $A$, and the state of the system after measurement can be expressed as ${{\rho}_{\hat{R}B}} = \sum_{i} {\left( {{{\left| {\psi _i} \right\rangle }^{\hat{R}}}\left\langle {\psi _i} \right| \otimes {\mathds{I}_B}} \right){{\rho}_{AB}}\left( {{{\left| {\psi _i} \right\rangle }^{\hat{R}}}\left\langle {\psi _i} \right| \otimes {\mathds{I}_B}} \right)}$. Recently, various improvement for EURs has been put forward  \cite{MAB,SW,AKP,MLH,MNB,TPP,LM,PJC,LR1,SZ,LR2,JZ,SL,YLX,FA,mf11,xbf,xbf1,mf,LLD,QHZ,TianYu}. For examples, Pati \emph {et al.} \cite{AKP} considered the optimized   EUR based on quantum discord,
and Adabi \emph {et al.} \cite{FA} strengthened  the lower bound of  EUR by considering    Holevo quantities. Furthermore, some promising experiments have been demonstrate  the proposed entropic uncertainty relations via all-optics and solid-state  platforms \cite{Rp,Cfl,ZY,wcm11,zxc,wml,hyw11,zxc11,wcxy1,Yong}.

Quantum systems exhibit many interesting traits within the framework of relativity, and had attracted much attention. The EURs of curved space-time have also been investigated by some authors \cite{JF,JL,DWW,FMDW}. {Actually, there are various curved space-time, including Schwarzschild space-time, Reissner-Nordstr$\mathrm{\ddot{o}}$m space-time, Garfinkle-Horowitz-Strominger space-time \cite{GWG,GKM}, etc.
It is well known that, the Schwarzschild black hole is regarded as one of basic and important black hole models. It assumes that the black hole is a spherically symmetric celestial body without rotation and charge. In order to better understand quantum properties of the black hole, examining EUR and quantumness in Schwarzschild space-time is required.} To illustrate more clearly, we describe the EUR in the context of black holes using the uncertainty game. Assuming that the system is prepared in the state of $\rho_{ABC}$, and particles $A$, $B$, and $C$ are sent to Alice, Bob, and Charlie, respectively. While particles $B$ and $C$ plunges into the black hole and hovers closely to its event horizon, and particle $A$ is situated in a flat space-time. After measuring her particles with $Q$ or $R$, Alice notifies Bob and Charlie of her measurement choice. The aim is to estimate Alice's measurement results as precisely as they can. More generally, multi-observable EURs in the many-body systems have been examined. For examples, some authors { \cite{LLD,QHZ,TianYu}} have novelly derived EURs with  $n$ measurements in $n+1$-party  systems. Herein, we derived a new and universal EUR with arbitrary measurement, and specifically observed the properties of quantum measured uncertainty and coherence in the background of {Schwarzschild} space-time.

The Letter is organized as follows: in Sec. \ref{sec2}, we firstly present a new and universal EUR for $m-$measurement within a $(n+1)-$party framework. In Sec. \ref{sec3},   the vacuum structure of {the Schwarzschild} black hole is reviewed briefly. In Sec. \ref{sec4}, we investigate the intrinsic relationship among the entropic uncertainty,  quantum coherence and mutual information in  {Schwarzschild} black holes. Finally, we end up with a concise  summary.

\section{ENTROPIC UNCERTAINTY RELATION FOR ARBITRARY MULTIPARTITE AND MULTI-OBSERVABLE
}
\label{sec2}
In this section, we first derive a new and universal EUR for arbitrary multiple measurements in multipartite systems, and then we present a stronger EUR based on considering the Holevo quantity.

{\bf Theorem 1.}\label{Th1}
By considering all possible measurements in arbitrary multipartite system, we can derive a simply constructed entropic uncertainty relation in the framework of multi-measurements:
\begin{equation}
\begin{split}
\sum_{j=1}^{n}\sum_{i=1}^{m} S(\hat{M}_i|B_j)\ge & -\frac{n}{m-1} \sum_{i\ne k,i=1}^{m} \log_2c_{ik} \\
&+\frac{m}{2}\sum_{j=1}^{n}S(A|B_j) ,
\label{Eq.2.2}
\end{split}
\end{equation}
where $m,n\in N^+$, $m$ represents the number of observable, while $n$ denotes the number of quantum memories.

{\bf Proof.} Suppose there are $m-$measurement in $n-$party systems. Herein we can list the conditional entropies for all possible measurements as {below}
\begin{equation}
\begin{split}
S(\hat{M}_1|B_1)\ \ S(\hat{M}_2|B_1)\ \  &S(\hat{M}_3|B_1)\  ... \ S(\hat{M}_m|B_1) \\
S(\hat{M}_1|B_2)\ \ S(\hat{M}_2|B_2)\ \  &S(\hat{M}_3|B_2)\ ...  \ S(\hat{M}_m|B_2) \\
S(\hat{M}_1|B_3)\ \ S(\hat{M}_2|B_3)\ \  &S(\hat{M}_3|B_3)\  ... \ S(\hat{M}_m|B_3) \\
&\vdots \\
S(\hat{M}_1|B_n)\ \ S(\hat{M}_2|B_n)\ \ &S(\hat{M}_3|B_n)\  ... \ S(\hat{M}_m|B_n)
\label{Eq.2.1}
\end{split}
\end{equation}
where, $\hat{M}_i\ (i=1,\cdots,m)$ is denoted as the observable to be measured, and $B_j\ (j=1,\cdots,n)$ represents quantum memory.

Firstly, we consider all the possible measurement pairs $(\hat{M}_i, \hat{M}_j)$ and particle $B_1$ as quantum memory in first row in Eq. (\ref {Eq.2.1}), and resort to the inequality in Eq. (\ref {Eq.1.4}),
we can obtain
\begin{equation}
\begin{split}
\sum_{i=1}^{m} S(\hat{M}_i|B_1)\ge -\frac{1}{m-1} \sum_{i\ne k,i=1}^{m} \log_2c_{ik}+\frac{m}{2}S(A|B_1) .
\label{Eq.2.3}
\end{split}
\end{equation}
In the same vein, we can also obtain the EUR outcomes of the  other rows. By adding up all the inequalities, we can achieve the construction of the desired Eq. (\ref{Eq.2.2}). \vskip 0.3cm

{\bf Theorem 2.}\label{Th2} By considering the mutual information and Holevo quantity and combining the previous result (\ref{Eq.2.2}), a new and tighter entropic uncertainty relation within
multiple measurements can be {obtained} as:
\begin{equation}
\begin{split}
\sum_{j=1}^{n}\sum_{i=1}^{m} S(\hat{M}_i|B_j)\ge &-\frac{n}{m-1} \sum_{i\ne k,i=1}^{m} \log_2c_{ik}\\
&+\frac{m}{2}\sum_{j=1}^{n}S(A|B_j)+ \max\{0,\Delta \}
\label{Eq.2.4}
\end{split}
\end{equation}
with $\Delta=\frac{m}{2} \sum_{j=1}^{n}\mathcal{I}(A:B_j)-\sum_{j=1}^{n} \sum_{i=1}^{m} \mathcal{H}(\hat{M}_i:B_j).$

{\bf Proof.} In the absence of quantum storage $B$, the inequality in Eq. (\ref{Eq.2.2}) can be expressed as:
\begin{equation}
\begin{split}
\sum_{i=1}^{m} S(\hat{M}_i)\ge -\frac{1}{m-1} \sum_{i\ne k,i=1}^{m} \log_2c_{ik}+\frac{m}{2}S(A).
\label{Eq.2.6}
\end{split}
\end{equation}
Linking the relation between conditional entropy and Holevo quantity  $ S(\hat{M}) = S(\hat{M}|B)+\mathcal{H}(\hat{M}:B)$ ($\mathcal{H}$ denotes the Holevo quantity), the element $\sum_{i=1}^{m}S(\hat{M}_i)$ of Eq. (\ref{Eq.2.6}) consequently can be written as:
\begin{equation}
\begin{split}
\sum_{i=1}^{m} S(\hat{M}_i) = \frac{1}{n} \left[\sum_{j=1}^{n}\sum_{i=1}^{m} S(\hat{M}_i|B_j) + \sum_{j=1}^{n}\sum_{i=1}^{m} \mathcal{H}(\hat{M}_i:B_j)\right].
\label{Eq.2.7}
\end{split}
\end{equation}
By substituting the above equation into the Eq. (\ref{Eq.2.6}), we can obtain the following inequality
\begin{equation}
\begin{split}
\sum_{j=1}^{n}\sum_{i=1}^{m} S(\hat{M}_i|B_j)\ge -\frac{n}{m-1} \sum_{i\ne k,i=1}^{m} \log_2c_{ik}\\
+\frac{mn}{2} S(A)
 -\sum_{j=1}^{n} \sum_{i=1}^{m} \mathcal{H}(\hat{M}_i:B_j).
\label{Eq.2.8}
\end{split}
\end{equation}
Because of the mutual information can written as the form of $\mathcal{I}(A:B) = S(A) - S(A|B)$. Substituting $S(A)=\mathcal{I}(A:B)+S(A|B)$ into the above formula (\ref{Eq.2.8}).
As a result, our new entropic uncertainty relation (i.e., Eq. (\ref{Eq.2.4})) can be obtained by comparing with  Eq. (\ref{Eq.2.2}) and taking the optimization over the derived results.
It is obvious that the lower bound of Eq. (\ref{Eq.2.4}) is tighter than that of Eq. (\ref{Eq.2.2}), verifying that the latter EUR outperforms the previous one.

\section{Vacuum structure of Dirac particles in the context of Schwarzschild black holes;}
\label{sec3}
We first go over the vacuum states of the {Schwarzschild} black hole. Generally, one {Schwarzschild} black hole can be described by
\begin{equation}
\begin{split}
{ds^{2} =} & {\left(1-\frac{2M}{r}\right) ^{-1}dr^{2} - \left(1-\frac{2M}{r}\right) dt^{2}}\\
&{+r^2(d\theta ^{2}+\sin^{2}\theta d\phi ^{2} ) ,}
\label{Eq.008}
\end{split}
\end{equation}
where {$M$ is the mass of the black hole.} Additionally, the Hawking temperature can be expressed as {$T = \frac{1}{8 \pi M}$} \cite{RKRB}. For the sake of  simplification, we set the gravitational constant $G$, the reduced Planck constant $\hbar $, the speed of light $c$, and the Boltzmann constant $k_B$ to unit $1$. In this case, Dirac's equation can be written as:
\begin{equation}
\begin{split}
\left [ \gamma ^{a} e_{a}^{\mu} \left ( \partial _{\mu } +\Gamma _{\mu }  \right ) \right ] \psi =0,
\label{Eq.009}
\end{split}
\end{equation}
where $\gamma ^{a}$ stands for Dirac matrices and $\Gamma _{\mu }$ is the spin connection coefficient. A set of solutions for positive frequency waves can be obtained by solving the above equation:
\begin{numcases}{}
 \psi _{k}^{I+} = \zeta e^{-i\omega \mu },
  \label{Eq.010}\\
 \psi _{k}^{II+} = \zeta e^{i\omega \mu }.
\label{Eq.011}
\end{numcases}
They respectively represent the solutions with mode $k$ in the regions of outside and inside the event horizon, and $\zeta$ is a four-component Dirac spinor. $\mu = t - r_*$ stands for retarded time and $\nu = t + r_*$ stands for advanced time, where {$r_* = r + 2 M \ln [ \frac{r-2M}{2 M} ]$} represents the tortoise coordinate. By taking the above two positive-frequency wave solutions as a set of completely orthogonal basis, and expanding the Dirac field, one can obtain
\begin{equation}
\begin{split}
\psi_{out} = \sum\limits_{\chi = I , II }\int dk(\alpha_{k}^{\chi} \psi_{k}^{\chi +})(\beta_{k}^{\chi *} \psi_{k}^{\chi -}),
\label{Eq.012}
\end{split}
\end{equation}
where $\alpha$ and $\beta$ are the fermion's annihilation operator and antifermion's creation operator respectively. According to the relation between Kruskal coordinates and black hole coordinates, a new set of orthogonal basis can be obtained:
\begin{numcases}{}
 {\phi_{k}^{I+} = e^{2 \pi M \omega_k }  \psi _{k}^{I+} + e^{-2 \pi M \omega_k }  \psi _{-k}^{II-}}
  \label{Eq.013}\\
 {\phi_{k}^{II+} = e^{-2 \pi M \omega_k }  \psi _{-k}^{I-} + e^{2 \pi M \omega_k }  \psi _{k}^{II+}.}
\label{Eq.014}
\end{numcases}
In this way, the new representation of the Dirac field in Kruskal space-time is
\begin{equation}
\begin{split}
&\psi_ {out} = \\
&{\sum\limits_{\chi =I , II } \int dk\frac{1}{\sqrt{2\cosh(4 \pi M \omega_k )} }  (a _{k}^{\chi} \phi _{k}^{\chi +} + b _{k}^{\chi *} \phi _{k}^{\chi -}),}
\label{Eq.015}
\end{split}
\end{equation}
where $a_{k}^{\chi}$ and $b_{k}^{\chi *}$ are the fermion annihilation and antifermion creation operators acting on the Kruskal vacuum respectively. By Bogoliubov transform, based on Eqs. (\ref{Eq.012}) and (\ref{Eq.015}), we can get the annihilation operator in the following form
\begin{equation}
\begin{split}
a_{k}^{I} = u\alpha _{k}^{I}-v\beta  _{k}^{II*},
\label{Eq.016}
\end{split}
\end{equation}
 {where, $u = (e^{-8 \pi M \omega_k}+1)^{-\frac{1}{2} }$ and $v = (e^{8 \pi M \omega_k}+1)^{-\frac{1}{2} }$.}

\begin{figure}
\centering
\includegraphics[width=8cm]{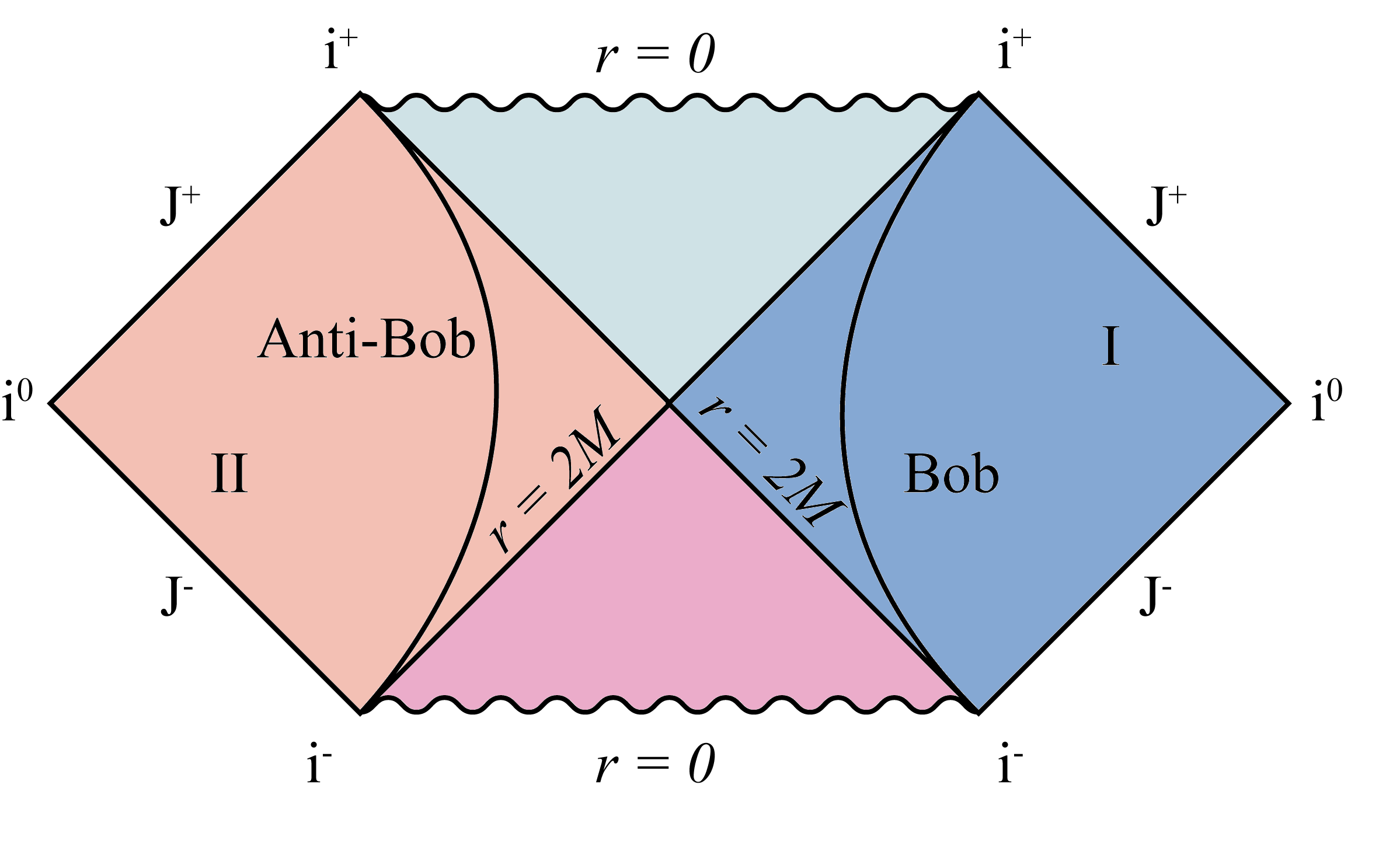}
\caption{The Penrose diagram of the {Schwarzschild} space-time indicating Bob and Anti-Bob trajectories. $i^0$ denotes the spatial infinities, $i^+$ and $i^-$ represents timelike future and past infinity. $J^+$ and $J^-$ shows lightlike future and past infinity, respectively. $r = 2M$ denotes the event horizons and {$r = 0$} represents the singularity of the black hole.
}
\label{Penrose diagrams}
\end{figure}

The Penrose diagram of {Schwarzschild} space-time has been sketched as  Fig. \ref{Penrose diagrams}. Since the physically accessible region $I$ and the physically inaccessible region $II$ are causally disconnected, {after some calculations for the Bogoliubov transformation, the vacuum state and the only excited state of the $k$-mode Kruskal particle can be expressed as}
\begin{equation}
\begin{split}
\left | 0 \right \rangle _k ^+ = u\left | 0 \right\rangle_{I}^+ \left | 0 \right\rangle_{II}^- + v\left | 1 \right\rangle_{I}^+ \left | 1 \right\rangle_{II}^-
\label{Eq.017}
\end{split}
\end{equation}
and
\begin{equation}
\begin{split}
\left | 1 \right \rangle _k ^+ = \left | 1\right \rangle _{I} ^+ \left | 0 \right \rangle _{II} ^- .
\label{Eq.018}
\end{split}
\end{equation}
The superscripts ``$+$'' and ``$-$'' in the expressions of these two states stand for particles and antiparticles, respectively, and the subscripts $I$ and $II$ stand for the outside and inside of the horizon.
\begin{figure}[htbp]
\centering
\subfigure[]{
  \includegraphics[width=0.225\textwidth]{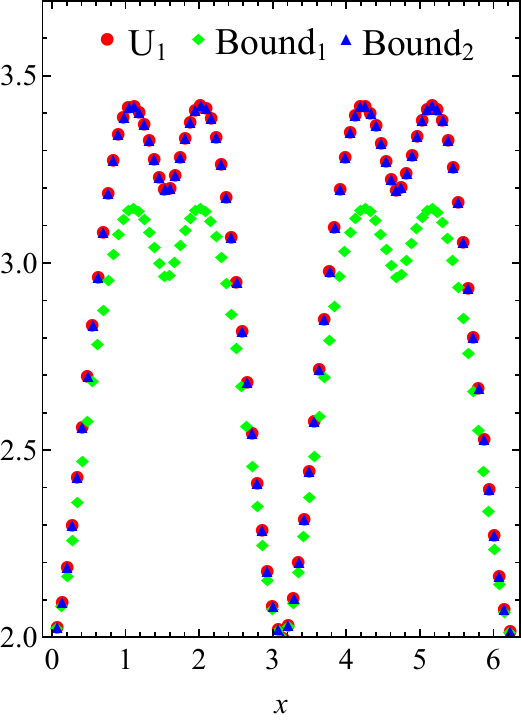}
}
\subfigure[]{
  \includegraphics[width=0.23\textwidth]{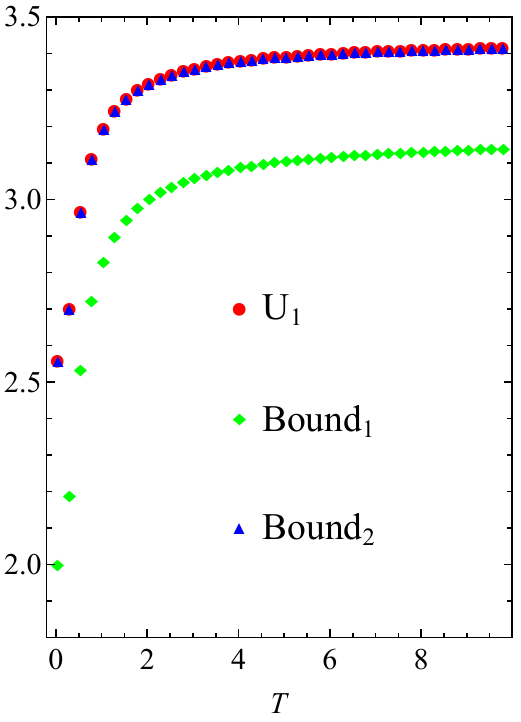}
}
\caption{Entropic uncertainty relations  for two-measurement ($\hat{\sigma_x}$, $\hat{\sigma_z}$) under the W-like initial state $\rho^{\rm W} _{AB_{I}C_{I}}$ in {Schwarzschild} black hole. The red circle represents the entropic uncertainty, the green diamond represents $\mathrm{Bound_1}$ in Eq. (\ref{Eq.4.2}) and the blue triangle represents $\mathrm{Bound_2}$ in Eq. (\ref{Eq.4.3}). Graph (a) shows uncertainty and the lower bounds vs the state's parameter $x$ with the { Hawking temperature $T = 10$} and state's parameter $y = \frac{\pi}{4}$, and Graph (b) shows uncertainty and the lower bounds versus {$T$} with $x = \frac{\pi}{3}$ and $y = \frac{\pi}{4}$.
}
\label{W-U}
\end{figure}

\section{Entropic uncertainty relation, coherence and mutual information in Schwarzschild black hole}
\label{sec4}

In order to examine the connection between the entropic uncertainty relation, quantum coherence, and mutual information in the context of {Schwarzschild} black holes, a three-qubit system with state of $\rho_{ABC}$ is employed as the initial state. To be explicit, particles $A$, $B$, and $C$ are located in the asymptotically flat region outside the event horizon of the black hole. Then,  particles $B$ and $C$ fall toward the {Schwarzschild} black hole and hover around its event horizon, while particle $A$ stays in the asymptotically flat region. Consequently, the  system's state can be reexpressed as $\rho_{A {B_{I}} {B_{II}} {C_{I}} {C_{II}} }$. We focus on obtaining the state $\rho _{AB_{I}C_{I}}$ in the physically accessible region by tracing the $B_{II}C_{II}$ in the quantum memory, since the two regions inside and outside the event horizon are disconnected.
In addition,  the two-dimensional Pauli operators are chosen as the incompatible measurements. We have $q_{MU} = - \log_2 c \left ( \hat{R},\hat{Q} \right ) = 1 $ with $\hat{R}$ and $\hat{Q} \in \{ \hat{\sigma_x}, \hat{\sigma_y}, \hat{\sigma _z} \} $, due to that the Pauli operators are typical mutual unbiased bases.

With regard to two measurements,  we can resort to $\hat{\sigma_x}$ and $\hat{\sigma _z}$  as the measurement operators. As a result,   our proposed  EURs
can be given by
\begin{equation}
\begin{split}
{U_{1}} \ge 2+S(A|B)+S(A|C) =: \mathrm{Bound_1},
\label{Eq.4.2}
\end{split}
\end{equation}
and
\begin{equation}
\begin{split}
{U_{1}} \ge 2+S(A|B)+S(A|C)+\max\{0,\Delta_1 \} =:\mathrm{Bound_2}
\label{Eq.4.3}
\end{split}
\end{equation}
where the uncertainty is denoted as ${U_{1}} := S(\hat{\sigma_x} | B )+S(\hat{\sigma_x} | C )+S(\hat{\sigma_z} | B )+S(\hat{\sigma_z} | C )$ and $\Delta_1 = \mathcal{I}(A:B)+\mathcal{I}(A:C) -\mathcal{H}(\hat{\sigma_x}:B) - \mathcal{H}(\hat{\sigma_x}:C) - \mathcal{H}(\hat{\sigma_z}:B)- \mathcal{H}(\hat{\sigma_z}:C).$

In addition,  $\hat{\sigma_x}$, $\hat{\sigma_y}$  and $\hat{\sigma _z}$ can be employed as  the measured observables in three-measurement scenario. Thereby, the proposed uncertainty relations can be written as:
\begin{equation}
{U_{2}} \ge 3 + \frac{3}{2} [ S(A|B)+S(A|C) ] =: \mathrm{Bound_3},
\label{Eq.4.6}
\end{equation}
and
\begin{equation}
\begin{split}
{U_{2}} \ge 3 + \frac{3}{2} [ S(A|B)+S(A|C) ]+\max\{0,\Delta_2 \} =:\mathrm{Bound_4}
\label{Eq.4.7}
\end{split}
\end{equation}
where  $
{U_{2}}= S(\hat{\sigma_x} | B )+S(\hat{\sigma_x} | C )+S(\hat{\sigma_y} | B )
+S(\hat{\sigma_y} | C )+S(\hat{\sigma_z} | B )+S(\hat{\sigma_z} | C )
$ is the amount of the uncertainty, and $\Delta_2 = \frac{3}{2}[ \mathcal{I}(A:B)+\mathcal{I}(A:C)] - \mathcal{H}(\hat{\sigma_x}:B)- \mathcal{H}(\hat{\sigma_x}:C) - \mathcal{H}(\hat{\sigma_y}:B)- \mathcal{H}(\hat{\sigma_y}:C)- \mathcal{H}(\hat{\sigma_z}:B)- \mathcal{H}(\hat{\sigma_z}:C)$. Besides, we set the frequency {$\omega_k=1$} in order to  simplify the calculation.

To verify our derived uncertainty relations (\ref{Eq.2.2}) and (\ref{Eq.2.4}), we proceed by taking into account one three-qubit W-like state as $\rho_{ABC}$, which is a unique kind of entangled state as:
\begin{equation}
\begin{split}
\rho^{\rm W} _{ABC} =  {| \mathrm{W} \rangle}{\langle \mathrm{W} | }
\label{Eq.4.9}
\end{split}
\end{equation}
with $| \mathrm{W}  \rangle =\cos x  | 001  \rangle  +\sin x\cos y | 010  \rangle
 +\sin x \sin y  | 100 \rangle$.

Fig. {\ref{W-U}}  depicts the entropic uncertainty and the two kinds of bounds (${\rm Bound}_1$ and ${\rm Bound}_2$) as functions of state's coefficient $x$ and {Hawking temperature $T$}, respectively. It is clear that our new form of entropic uncertainty relation is applicable to two measurements within three-qubit systems. Fig. {\ref{W-U}}(a) shows that the uncertainty and lower bound change periodically with the increase of the state parameter $x$. {The uncertainty increases with the growth of $T$ as shown in Fig. {\ref{W-U}}(b). Meanwhile,} the quantum coherence is greatly broken and the uncertainty increases to maximum, which will be further discussed later in this article. In addition, $\mathrm{Bound_2}$ is obviously tighter than $\mathrm{Bound_1}$  shown in Fig. {\ref{W-U}}. We can conclude: (1)
our proposed inequalities (\ref{Eq.2.2}) and (\ref{Eq.2.4}) are hold; (2) $\mathrm{Bound_2}\geqslant \mathrm{Bound_1}$ is satisfied all the time, which implies the lower bound of the latter outperforms  than that in the former.

\begin{figure*}[t]
\centering

  {\includegraphics[width=5.5cm]{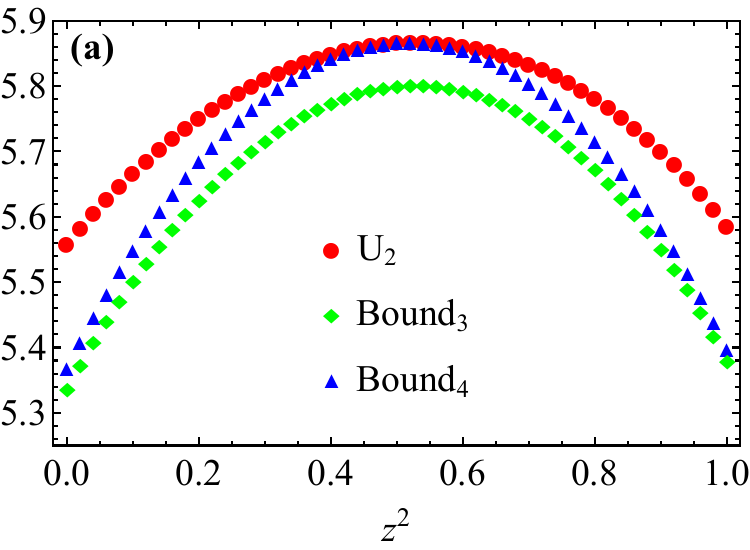} }\ \
  {\includegraphics[width=5.5cm]{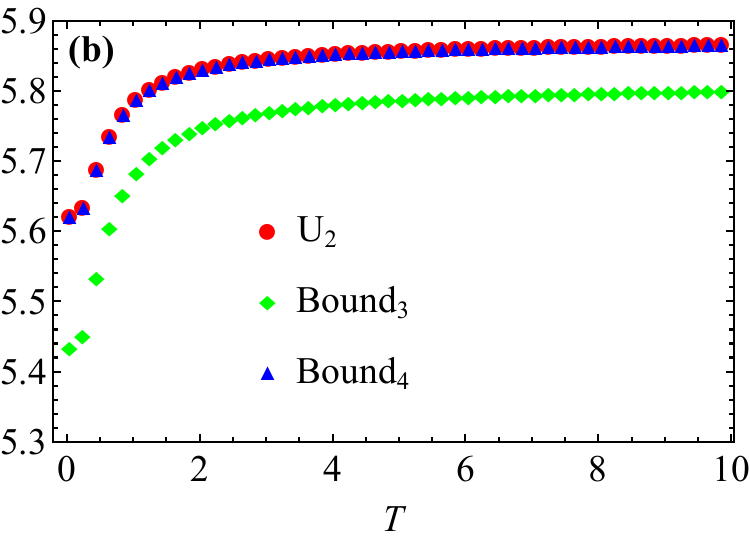} }\ \
  {\includegraphics[width=5.5cm]{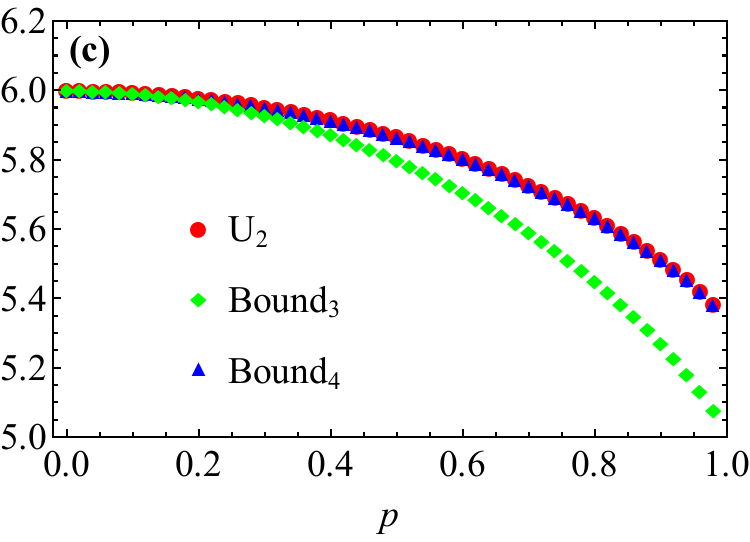} }

\caption{Entropic uncertainty relations of Werner state $\rho^{\rm{Werner}} _{AB_{I}C_{I}}$ in the {Schwarzschild} black hole. The Graphs show the case of three-measurement ($\hat{\sigma_x}$, $\hat{\sigma_y}$, $\hat{\sigma_z}$).  The red circle represents the entropic uncertainty, green diamond represents $\mathrm{Bound_3}$ in Eq. (\ref{Eq.4.6}) and blue triangle represents $\mathrm{Bound_4}$ in Eq. (\ref{Eq.4.7}). In Graph (a), uncertainty and the lower bounds vs $z^2$ with {$T=10$} and $p = 0.5$; in Graph (b), uncertainty and the lower bounds vs {$T$} with $p=0.5$ and $z^2 = 0.5$; in graph (c), uncertainty and the lower bounds vs the purity parameter $p$ with $z^2 = 0.5$ and {$T=10$}.
}
\label{Werner-U}
\end{figure*}

\begin{figure*}[htbp]
\subfigure[]
  {
      \label{C1}
      \includegraphics[width=5.5cm]{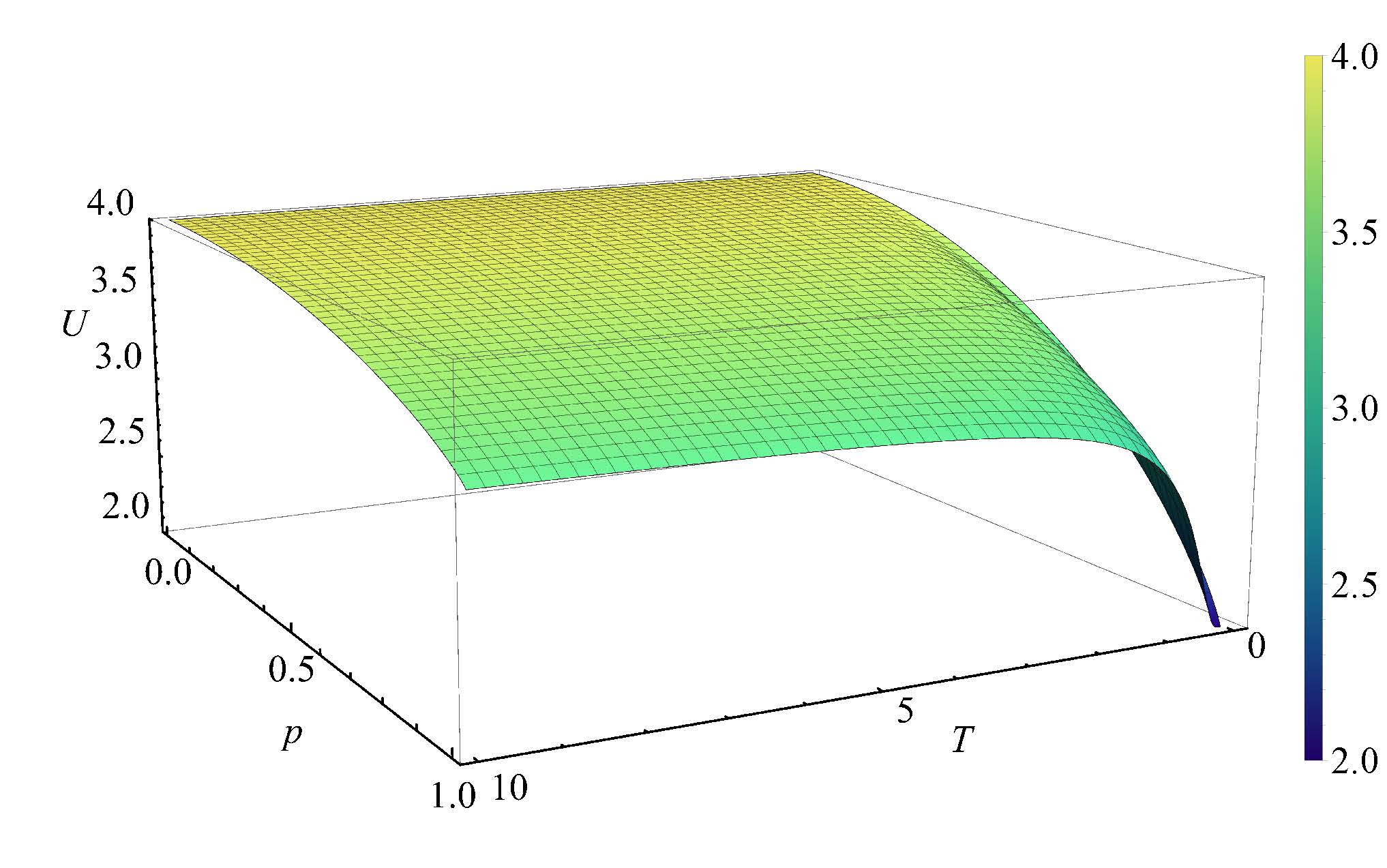}
  }
\subfigure[]
  {
      \label{C3}
      \includegraphics[width=5.5cm]{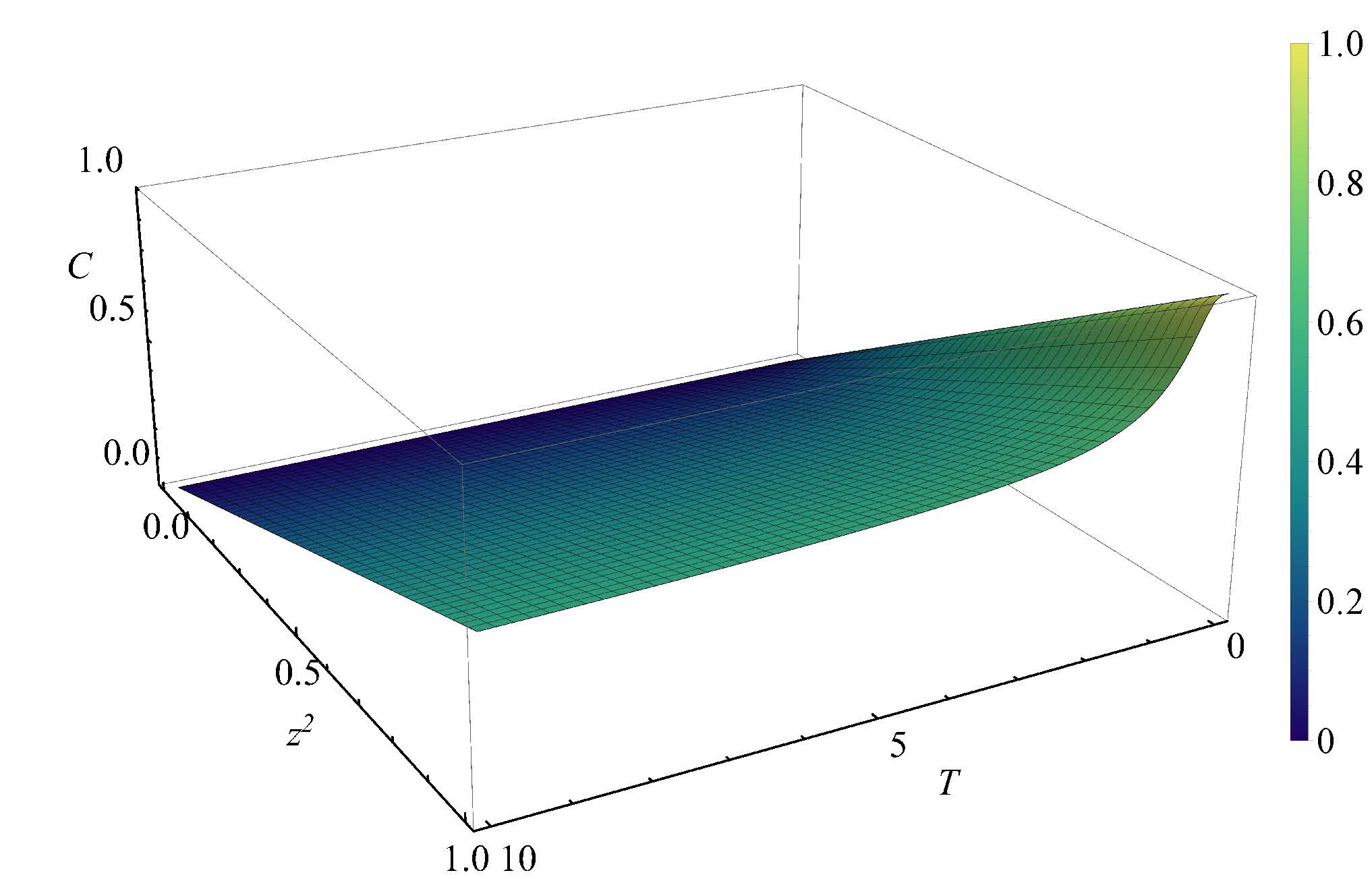}
  }
\subfigure[]
  {
      \label{C5}
      \includegraphics[width=5.5cm]{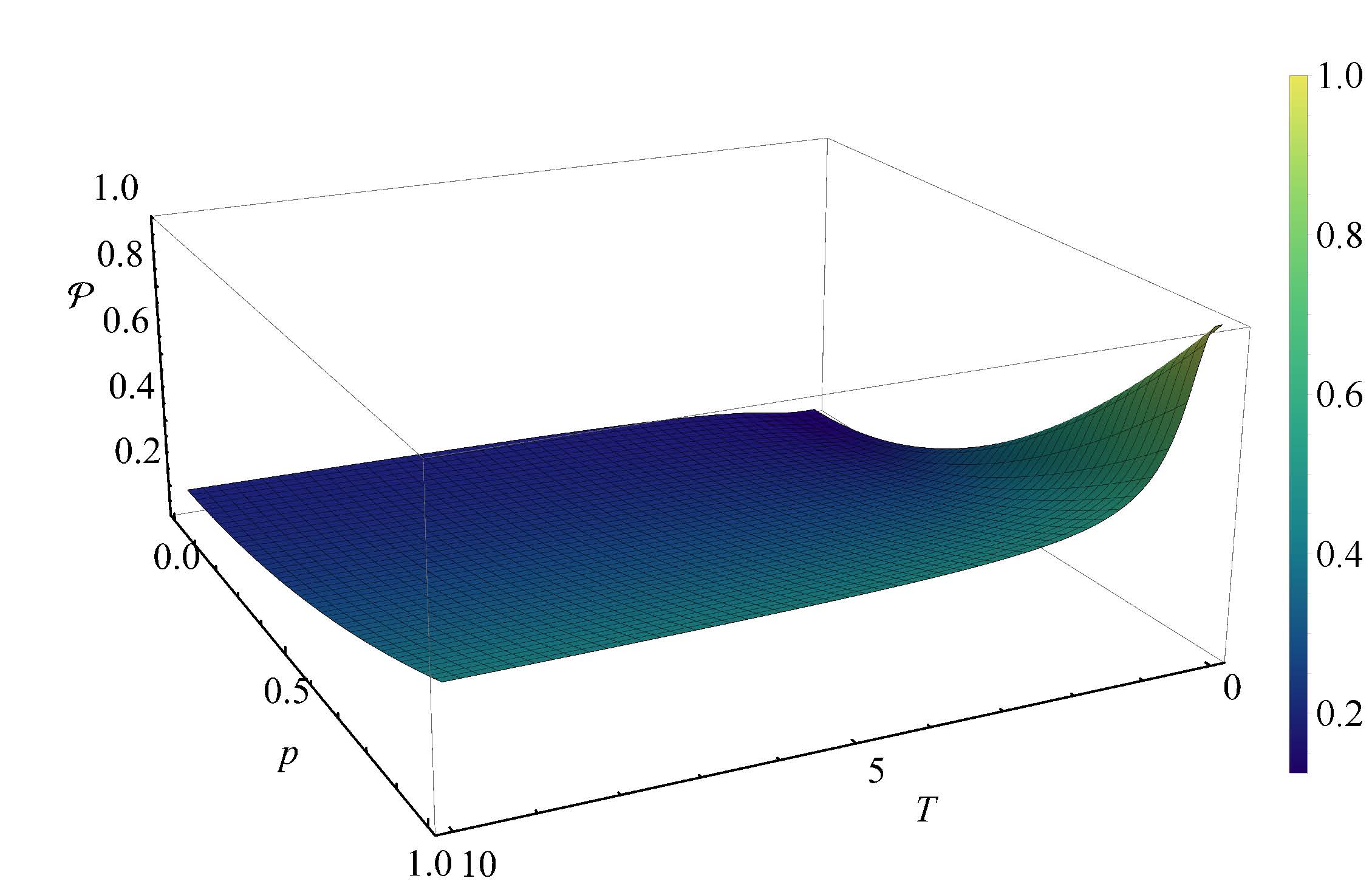}
  }

\subfigure[]
  {
      \label{C2}
      \includegraphics[width=5.5cm]{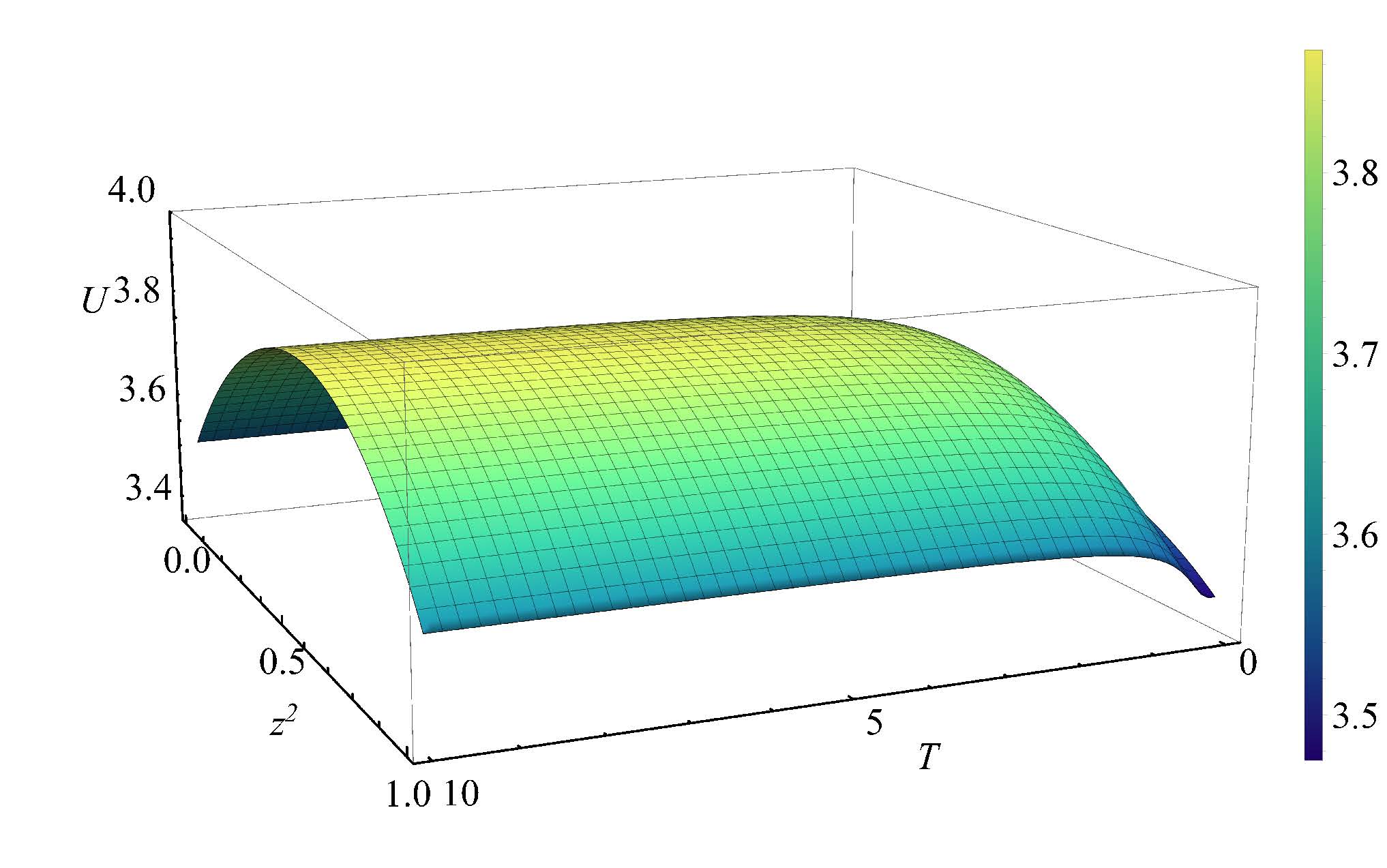}
  }
\subfigure[]
  {
      \label{C4}
      \includegraphics[width=5.5cm]{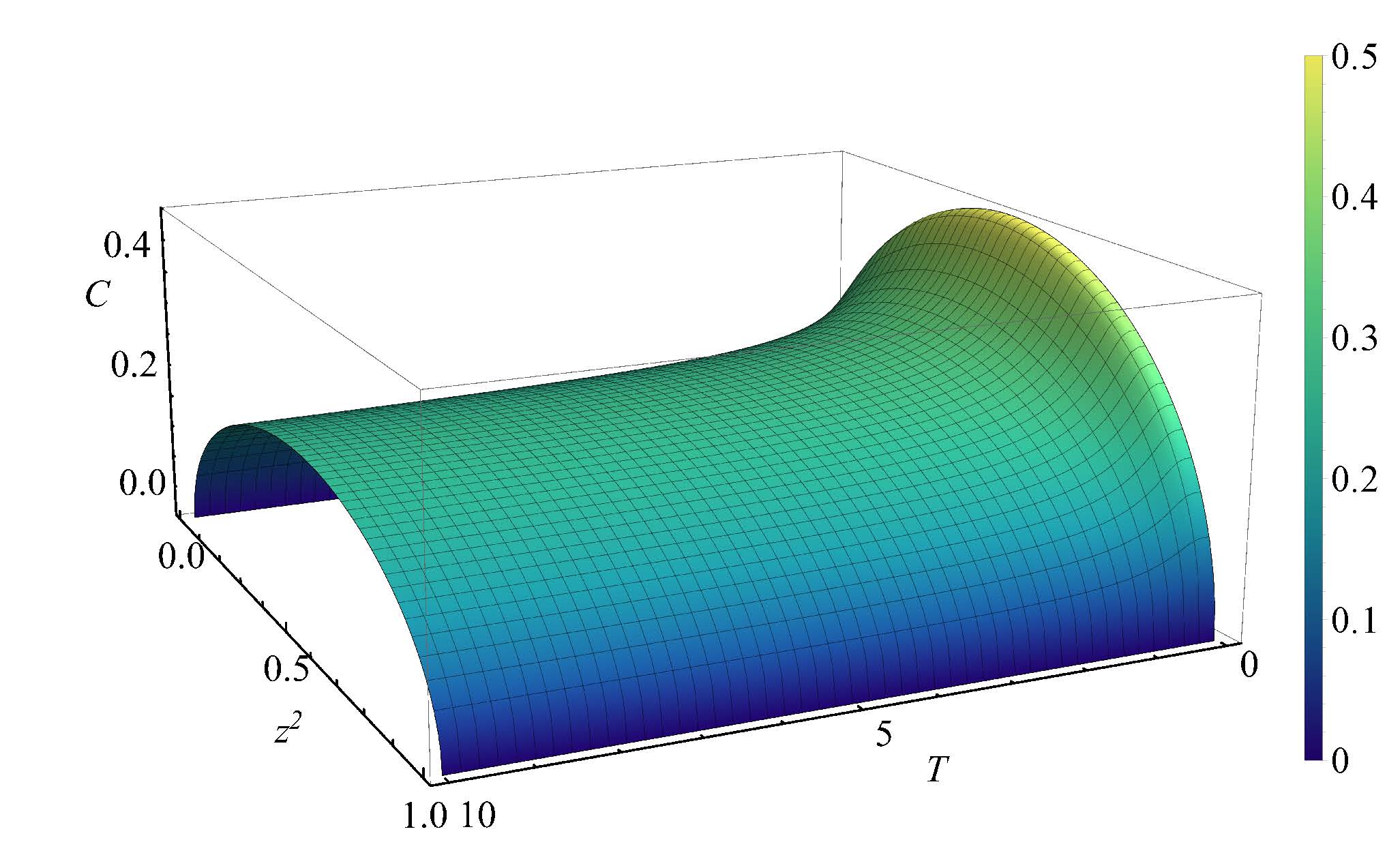}
  }
\subfigure[]
  {
      \label{C6}
      \includegraphics[width=5.5cm]{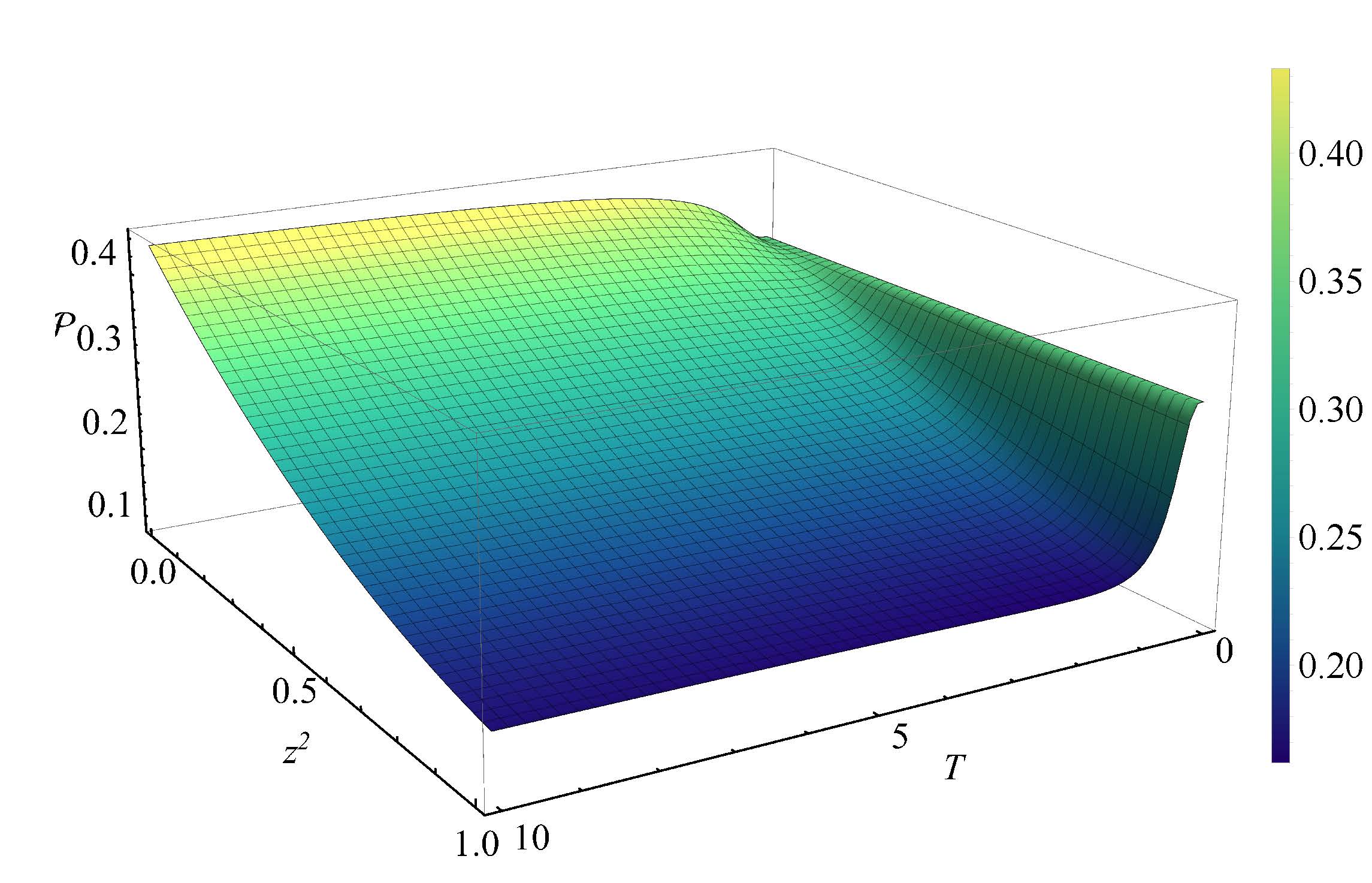}
  }
\caption{Entropic uncertainty $U$, coherence $C$ and purity $\cal P$ of Werner state $\rho^{\rm Werner} _{AB_{I}C_{I}}$ in the {Schwarzschild} black hole. Graphs (a), (b) and (c) show the uncertainty, coherence and purity versus {$T$} and $p$ with $z^2=0.5$; Graphs (d), (e) and (f) show the uncertainty, coherence and purity as functions of {$T$} and $z^2$ with $p=0.5$, respectively.}
\label{Werner-2D}
\end{figure*}

\begin{figure}[htbp]
\centering
\includegraphics[width=8cm]{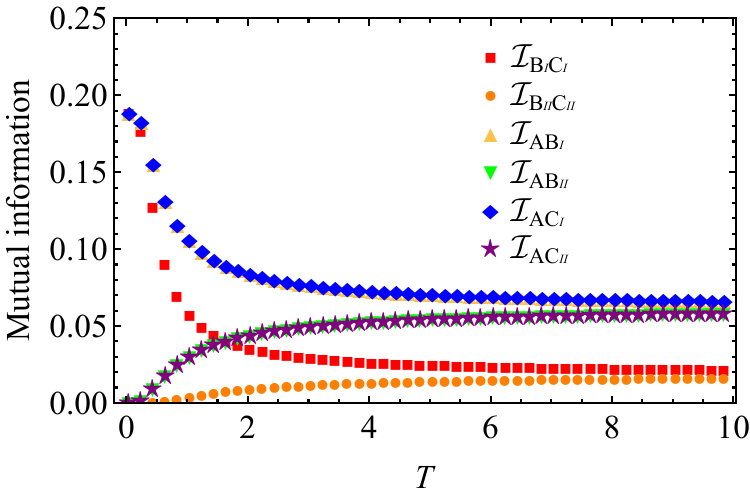}
\caption{ {Mutual information vs the {Hawking temperature $T$} for Werner state $\rho^{\rm Werner} _{AB_{I} B_{II} C_{I} C_{II}}$ as initial state in the {Schwarzschild} black hole with $p=0.5$ and $z^2 = 0.5$.} The red square, orange circle, yellow triangle, green inverted triangle, blue diamond and purple pentagram represent the mutual information of $B_{I}C_{I}$, $B_{II}C_{II}$, $AB_{I}$, $AB_{II}$, $AC_{I}$, and $AC_{II}$, respectively.}
\label{Werner-I}
\end{figure}

\begin{figure*}[htbp]

  {
      \includegraphics[width=5.5cm]{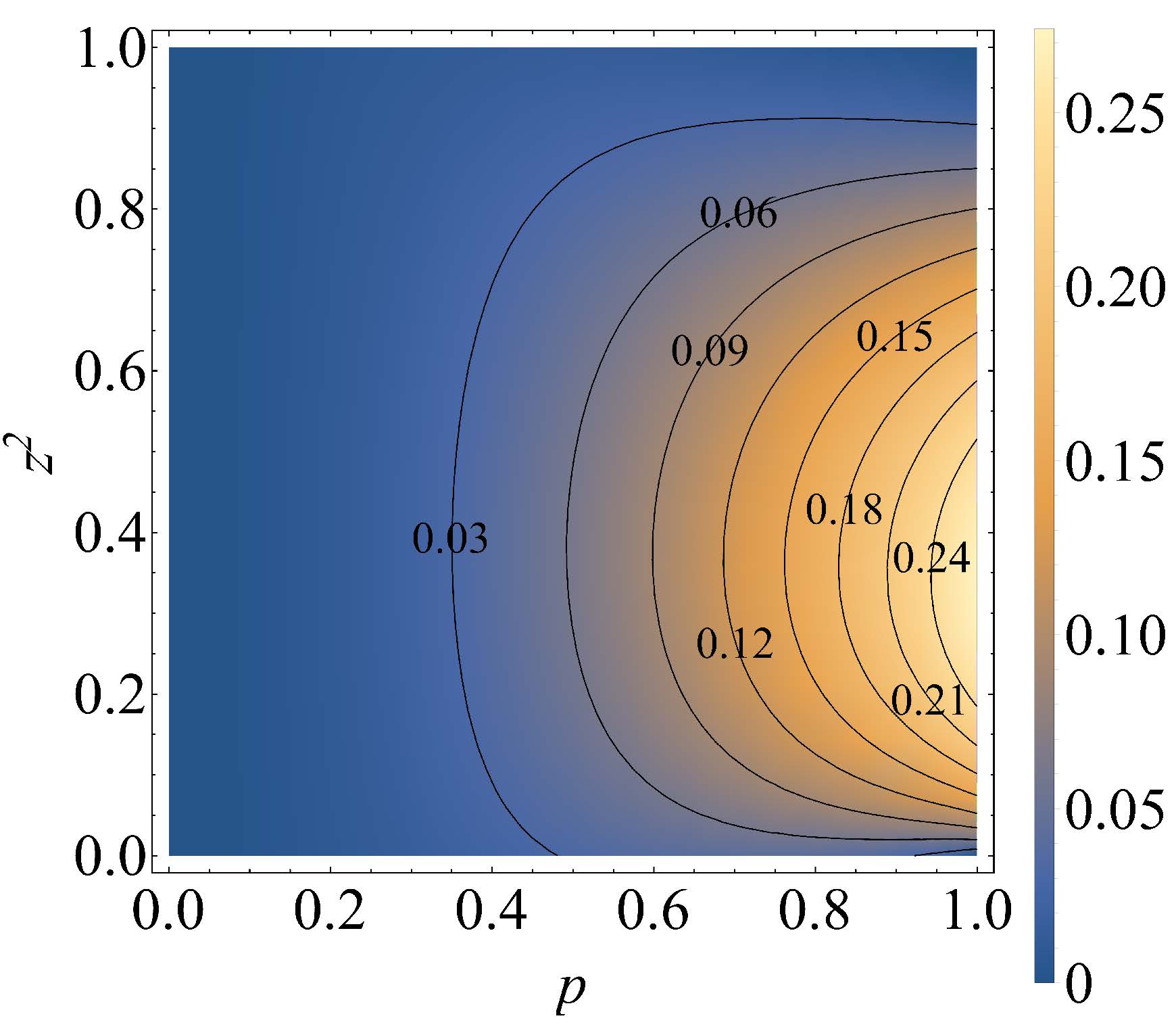}
  }
  {
      \includegraphics[width=5.5cm]{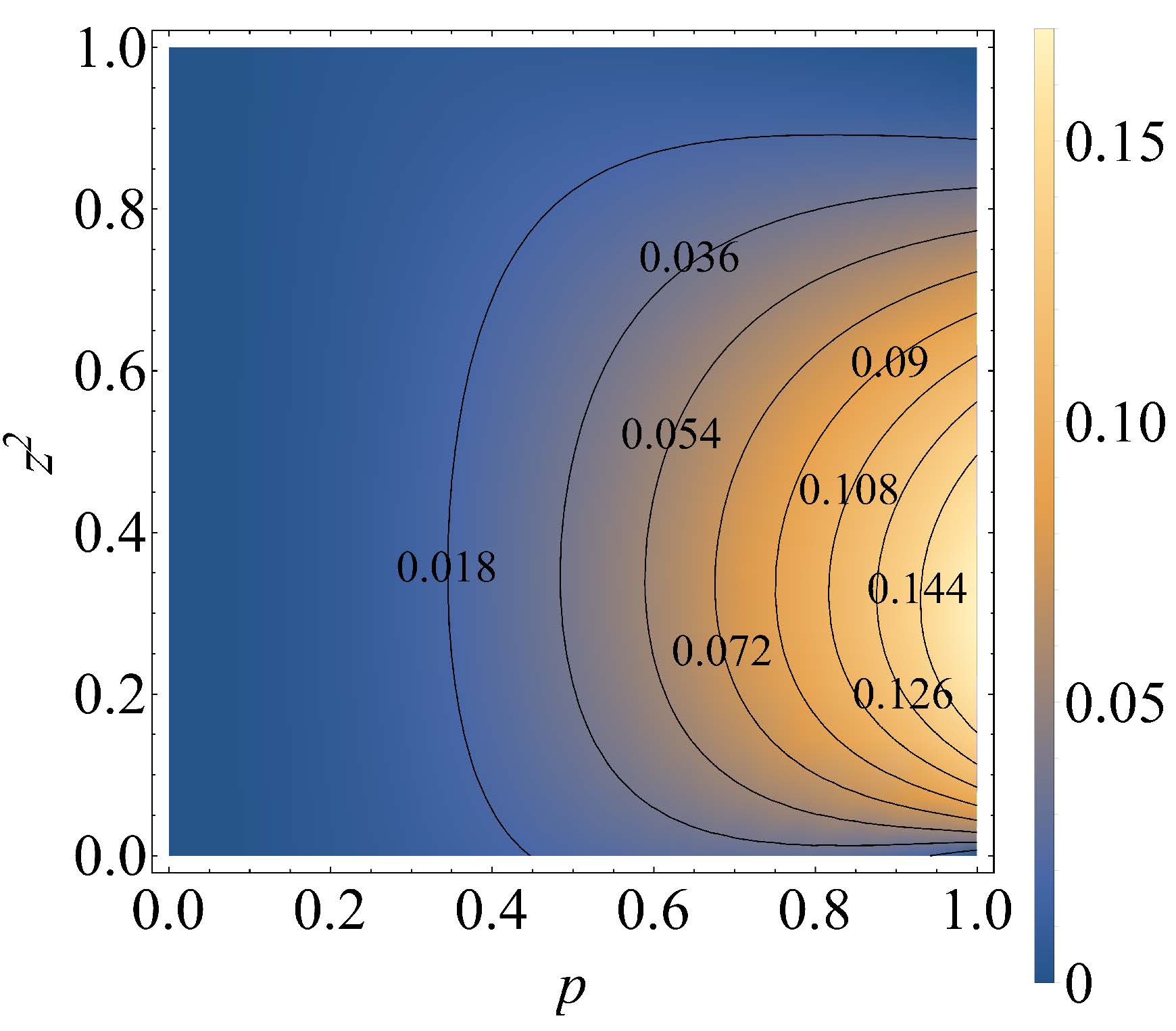}
  }
  {
      \includegraphics[width=5.5cm]{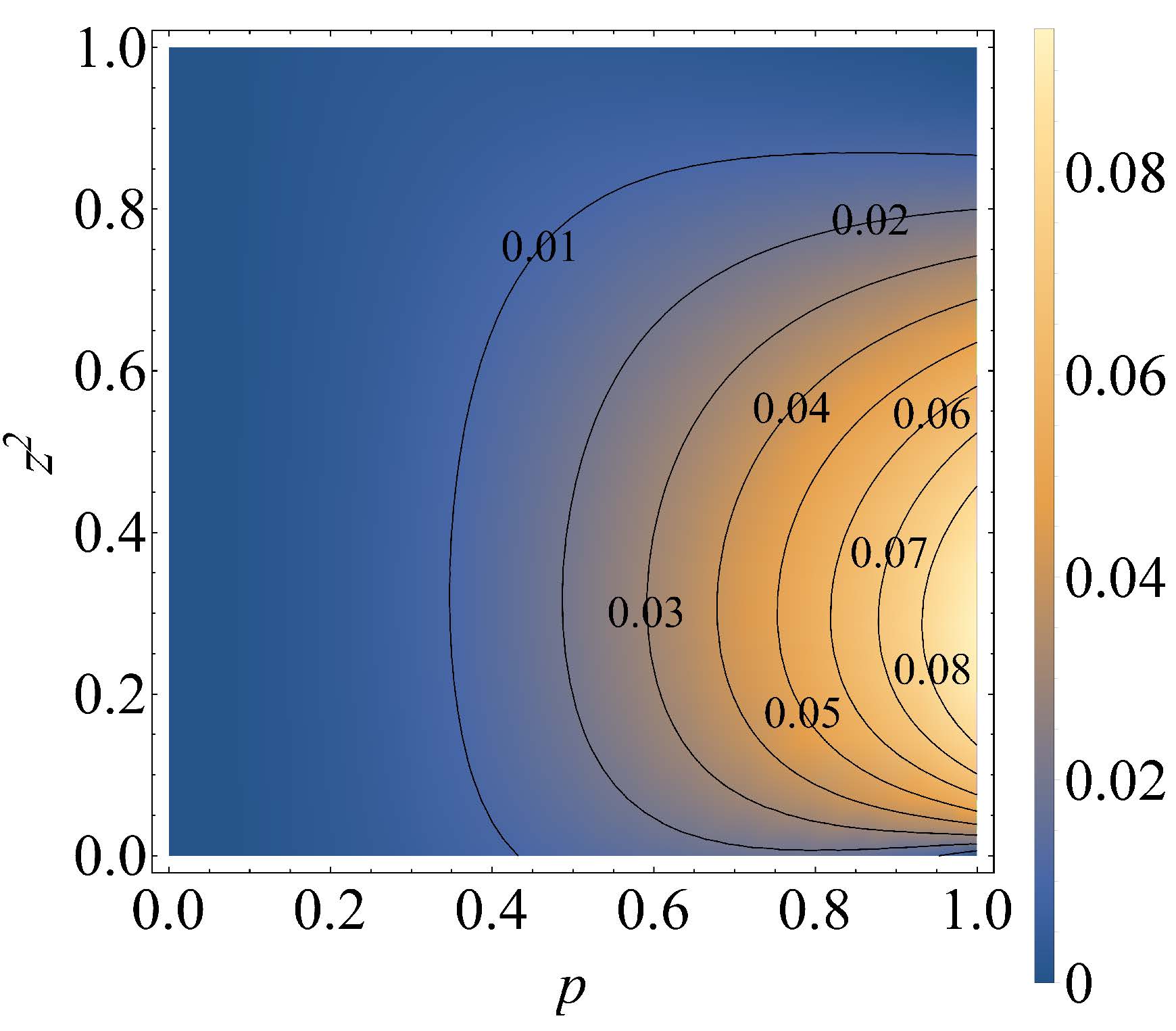}
  }

  {
      \includegraphics[width=5.5cm]{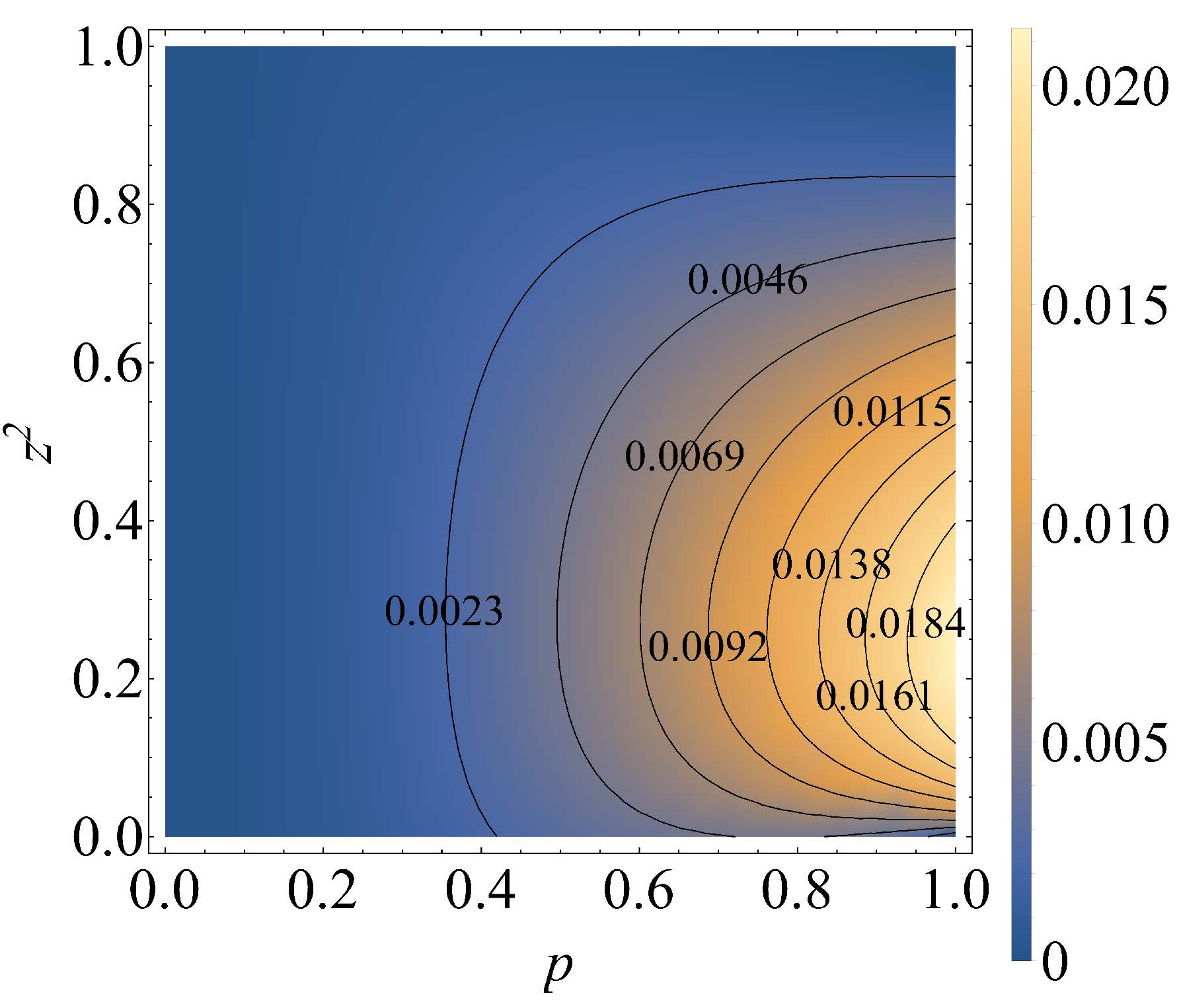}
  }
  {
      \includegraphics[width=5.5cm]{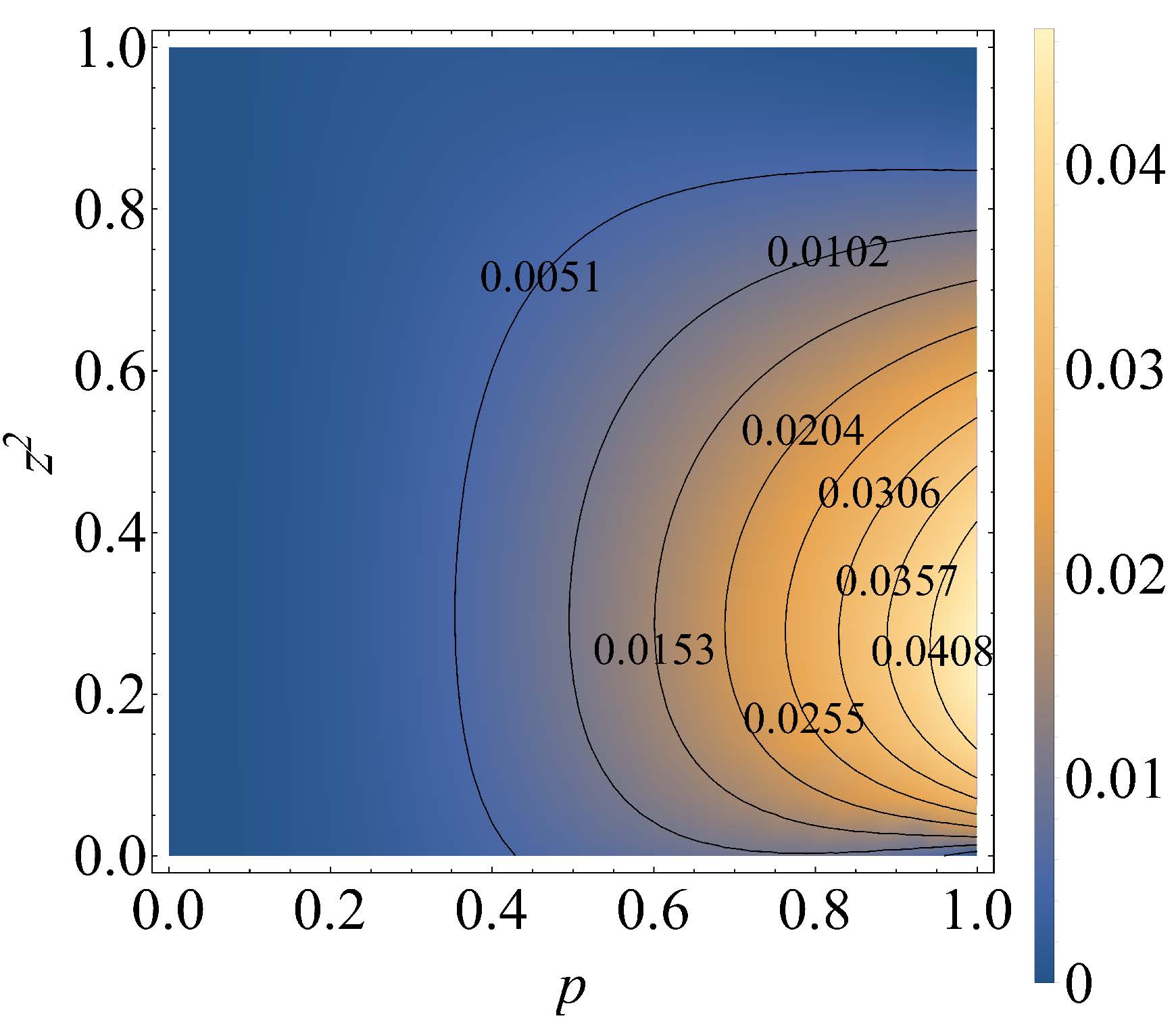}
  }
  {
      \includegraphics[width=5.5cm]{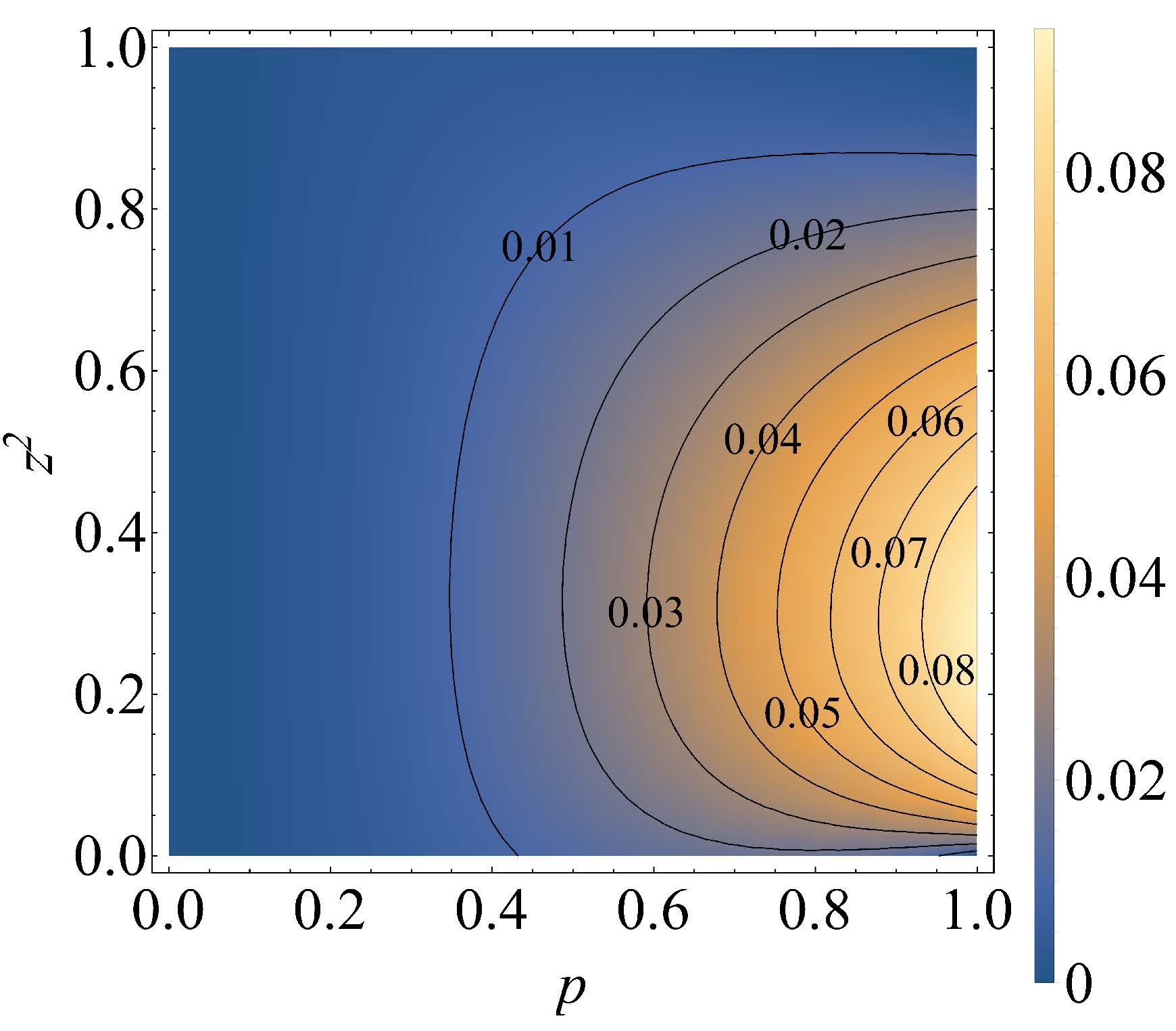}
  }
\caption{Mutual information as functions of the state's parameters $p$ and $z^2$ of Werner state $\rho^{\rm Werner} _{AB_{I}B_{II} C_{I}C_{II}}$ in the {Schwarzschild} black hole. Graphs (a), (b) and (c) plot ${\cal I}_{B_{I}C_{I}}$ with {$T = 1, 2, \infty$}. Graphs (d), (e) and (f) plot ${\cal I}_{B_{II}C_{II}}$ with {$T = 1, 2, \infty$},  respectively.}
\label{Werner-I-2D}
\end{figure*}

Without loss of generality, we also consider a three-qubit Werner state, which canonical is a type of mixed states:
\begin{equation}
\begin{split}
\rho ^{\rm Werner}_{ABC} = p {\left| \mathrm{GHZ} \right\rangle }{\left\langle \mathrm{GHZ} \right| } + \frac{(1-p)}{8} \mathbb{I}_{8 \times 8} ,
\label{Eq.4.11}
\end{split}
\end{equation}
with $| \mathrm{GHZ} \rangle = z | 000  \rangle + \sqrt{1-z^{2} }  | 111 \rangle$,  the purity of the system $p\in[0,1]$. The state is the maximally mixed state for $p=0$, while it becomes the Greenberger-Horne-Zeilinger-type (GHZ-like) state, a typical pure state, for $p=1$. Explicitly, entropic uncertainty and two bounds with respect to   the state parameter $z^2$, $p$ and {Hawking temperatur $T$} have been plotted in Fig. \ref{Werner-U}. As indicated in Fig. \ref{Werner-U}(a), the uncertainty increases first and then decreases with the growing $z^2$. Fig. \ref{Werner-U}(b) indicates that, near {$T = 0$}, the uncertainty increases rapidly, which result is similar to that in the W-like states. In Fig. \ref{Werner-U}(c), the uncertainty and lower bounds decrease monotonically as the purity $p$ grows, and the uncertainty will minimize as to $p=1$. What's more,  one can directly see that $\mathrm{Bound_4}\geqslant \mathrm{Bound_3}$ in terms of the current numerical analysis, which is essentially in agreement with the conclusion made before.

To better interpret the dynamics of the uncertainty in the background of {Schwarzschild} black hole, we will focus on examining coherence and purity of the system of interest. Technically, $l_1$-norm is used to quantify quantum coherence of a given system $\rho$ with the form of $C_{l_1}{(\rho)} = \sum_{i \ne j }\left |\rho _{i,j}  \right | $, and the purity of the system is denoted as ${\cal P} = \mathrm{Tr} (\rho^2)$. In this case, we can work out {$C_{l_1}{(\rho^{\rm Werner}_{AB_{I}C_{I}})} = \frac {2pz\sqrt{(1-z^{2})}} {1+e^{-\frac{\omega_k}{T}} }$} and the analytical expression of ${\cal P}$ is provided as {\hyperlink{appendixlink}{Appendix}}.  As shown in Figs. \ref{C1}, \ref{C3}, \ref{C2} and \ref{C4}, the uncertainty and coherence with respect to the {Hawking temperature $T$} always shows the opposite trend, indicating that there is an inverse correlation between uncertainty and quantum coherence regardless of the initial state's parameter $p$. In the physically accessible region, as illustrated by Fig. \ref{C5} , the system's purity is  mainly determined by the state's parameter $p$; the purity declines as {$T$ increases, which indicates that an increase in Hawking temperature $T$ could disrupt the purity of the system.} The purity with the growing {$T$} and $z^2$ has been plotted in Fig. \ref{C6}. Following the figure, as the Hawking temperature $T$ increases, the coherence $C$ of the system is gradually broken, resulting in a gradual decrease in purity $\cal P$. However, it is interesting to note that when $z^2$ is very small, which corresponds to that the initial state is the direct product states $| 111 \rangle$, the purity of the system will increase.
This manifests that the existence of excited states will prevent the coherence of the system from being destroyed due to the increase of $T$, so the purity of the system will increase. Thereby, we can conclude: (1) an increasing Hawking temperature will break the coherence of the system and increase the entropic uncertainty; (2) the uncertainty and coherence exist an anti-correlation characteristic with {variation of Hawking temperature $T$}; (3) when the Hawking temperature $T$ is small, the purity of the state $\rho^{\rm Werner} _{AB_{I}C_{I}}$ in the physically accessible region is  dominated by the purity of the original state $\rho^{\rm Werner}_{ABC}$, however the state parameter $z^2$ will govern the system's purity  when the Hawking temperature is large enough.

To clarify the dynamical mechanics of the uncertainty in {Schwarzschild} black hole, we next turn to  mutual information, which is able to show the information distribution in the whole system. Fig. \ref{Werner-I}
has plotted  the mutual information related to the {Hawking temperature $T$}. It can be seen that the physically accessible region mutual information ${\cal I}_{B_{I}C_{I}}$, ${\cal I}_{AB_{I}} $ and ${\cal I}_{AC_{I}}$ decrease {as $T$ increases}, while the physically inaccessible region ones increase.  This demonstrates the flow of information from physically accessible region to physically inaccessible region. Linking with the dynamics of uncertainty in Fig. \ref{Werner-U}(b), it is evident that the mutual information and uncertainty exhibit fully opposite trends as the Hawking temperature increases. Therefore, it is inferred that the Hawking effect results in reduction of the correlation within the physically accessible region, and consequently leads to inflation of the uncertainty.
Particularly, when {$T\rightarrow \infty$}, ${\cal I}_{B_{I}C_{I}}$ and ${\cal I}_{B_{II}C_{II}}$ tend to be equal, which indicates that when the Hawking temperature $T$ is large enough, the boundary between the physically accessible region and the physically inaccessible region becomes blurred. Mutual information $B_{I}, C_{I}$ and $B_{II}C_{II}$  with respect to the variation in $z^2$ and $p$ are plotted in Fig. \ref{Werner-I-2D}. From the figure, the mutual information  generally increases with the increase of purity $p$. As $T$ increases, the mutual information between $B_{I}$ and $C_{I}$ decreases, the mutual information between $B_{II}$ and $C_{II}$ increases, which indicates the information redistribution of the {Schwarzschild} black hole.

\section{Conclusions}
\label{sec6}
In summary, we have derived the generalized EUR for   $m$ measurements with $n+1$-party, and have further strengthened it via mutual information and Holevo quantity. The proposed EURs have been verified by utilization of three-qubit  W-like and Werner states in the background of {Schwarzschild} space-time.  Besides, we observed that Hawking radiation destroys quantum coherence and increases the uncertainty of the system. Furthermore, we revealed that the uncertainty of the system is anti-correlation with the system's purity. As the Hawking temperature increases, the mutual information of systems outside the black hole's event horizon decreases, while ones  within the event horizon increases. This suggests that the Hawking effect leads to the redistribution of information from a physically accessible region to a physically inaccessible region, which can be used to explain the uncertainty dynamics of the {Schwarzschild} black hole. Thus, it is believed that the current findings provide insight into the generalized uncertainty relation, and are beneficial to understand information paradox in the black holes.

\begin{acknowledgements}
This study was supported by the National Natural Science Foundation of China (Grant Nos. 12075001 and 61601002), and Anhui Provincial Key Research and Development Plan (Grant No. 2022b13020004).
\end{acknowledgements}

\hypertarget{appendixlink}{}
\appendix
\section*{APPENDIX}
The specific form of purity is given by:
\begin{equation}
\begin{split}
{\cal P} = &\frac{1}{64} \left[ u^8 (-1 + p)^2 + u^8 (-1 + p - 8 p z^2)^2 \right] \\
 &+ \frac{1}{64} \left[ -1 + v^2 (2 + v^2) (-1 + p) - 7 p + 8 p z^2 \right]^2 \\
 &+ \frac{1}{64} \left[ (1 + v^2)^2 (-1 + p) - 8 v^4 p z^2 \right]^2 \\
 &+ \frac{1}{32} u^4 \left\{ 1 - p + v^2 [1 +(-1 + 8 z^2)] \right\}^2 \\
 &+ \frac{1}{32} u^4 (1 + v^2)^2 (-1 + p)^2 - 2 u^4 p^2 z^2 (-1 + z^2)
\end{split}\label{appendix}
\end{equation}
with {$u = [e^{-\omega_k / T}+1]^{-\frac{1}{2} }$ and $v = [e^{ \omega_k / T}+1]^{-\frac{1}{2} }$.}


\begin{references}

\bibitem {WH} W. Heisenberg,
\href {https://doi.org/10.1007/BF01397280} {Z. Phys. {\bf 43}, 172 (1927).}

\bibitem {EHK} E. H. Kennard,
\href {https://doi.org/10.1007/BF01391200} {Z. Phys. {\bf 44}, 326 (1927)}.

\bibitem {HPR}	H. P. Robertson,  
\href {https://doi.org/10.1103/PhysRev.34.163} {Phys. Rev. {\bf 34}, 163 (1929)}.

\bibitem {HE} H. Everett, 
\href {https://doi.org/10.1103/RevModPhys.29.454}  {Rev. Mod. Phys. {\bf29}, 454 (1957)}.

\bibitem {IIH} I. I. Hirschman, 
\href {https://doi.org/10.2307/2372390}  {Am. J. Math. {\bf79}, 152 (1957)}.

\bibitem {DD} D. Deutsch,  
\href {https://doi.org/10.1103/PhysRevLett.50.631}{       Phys. Rev. Lett. {\bf 50}, 631 (1983)}.

\bibitem {KK} 	K. Kraus,  
\href {https://doi.org/10.1103/PhysRevD.35.3070}{       Phys. Rev. D {\bf 35}, 3070 (1987)}.

\bibitem {HM} 	H. Maassen and J. B. M. Uffink,  
\href {https://doi.org/10.1103/PhysRevLett.60.1103}{       Phys. Rev. Lett. {\bf 60}, 1103 (1988)}.

\bibitem {JMR}   J. M. Renes and J. C. Boileau,
\href {https://doi.org/10.1103/PhysRevLett.103.020402}{      Phys. Rev. Lett. {\bf103}, 020402 (2009)}.

\bibitem {MB}  	M. Berta, M. Christandl, R. Colbeck, J. M. Renes, and R. Renner,
\href {https://doi.org/10.1038/nphys1734}{  Nat. Phys. {\bf 6}, 659 (2010).}

\bibitem {MAB} M. A. Ballester and S. Wehner,  
\href {https://doi.org/10.1103/PhysRevA.75.022319 }
{           Phys. Rev. A {\bf 75}, 022319 (2007)}.
\bibitem {SW} S. Wu, S. Yu, and K. M{\o}lmer, 
\href {https://doi.org/10.1103/PhysRevA.79.022104  }
{            Phys. Rev. A {\bf 79}, 022104 (2009)}.
\bibitem {AKP}	A. K. Pati, M. M. Wilde, A. R. Usha Devi, A. K. Rajagopal, and Sudha,  
\href {https://doi.org/10.1103/PhysRevA.86.042105  }
{           Phys. Rev. A {\bf 86}, 042105 (2012)}.

\bibitem {MLH} M. L. Hu and H. Fan, Phys.
\href {https://doi.org/10.1103/PhysRevA.87.022314   }
{          Phys. Rev. A  {\bf 87}, 022314 (2013)}.

\bibitem {MNB} M. N. Bera, R. Prabhu, A. Sen(De), U. Sen,
\href {https://doi.org/10.1103/PhysRevA.86.012319   }
{          Phys. Rev. A   {\bf86}, 012319 (2012)     }

\bibitem {TPP} T. Pramanik, P. Chowdhury, and A. S. Majumdar,  
\href {https://doi.org/10.1103/PhysRevLett.110.020402    }
{          Phys. Rev. Lett. {\bf 110}, 020402 (2013)}.

\bibitem {LM}	L. Maccone and A. K. Pati,  
\href {https://doi.org/10.1103/PhysRevLett.113.260401    }
{          Phys. Rev. Lett. {\bf 113}, 260401 (2014).}

\bibitem {PJC} P. J. Coles and M. Piani, 
 \href {https://doi.org/10.1103/PhysRevA.89.022112  }
{         Phys. Rev. A {\bf 89}, 022112 (2014)}.

\bibitem {LR1}	{\L}. Rudnicki, Z. Puchala, and K. \.{Z}yczkowski,  
 \href {https://doi.org/10.1103/PhysRevA.89.052115  }
{        Phys. Rev. A {\bf 89}, 052115 (2014)}.

\bibitem {SZ}	S. Zozor, G. M. Bosyk, and M. Portesi,  
\href {https://iopscience.iop.org/article/10.1088/1751-8113/47/49/495302/meta  }
{         J. Phys. A {\bf 47}, 495302 (2014)}.

\bibitem {LR2}	{\L}. Rudnicki,  
 \href {https://doi.org/10.1103/PhysRevA.91.032123  }
{        Phys. Rev. A {\bf 91}, 032123 (2015)}.

\bibitem {JZ} J. Zhang, Y. Zhang, and C. S. Yu,  
\href {https://doi.org/10.1038/srep11701  }
{        Sci. Rep. {\bf 5}, 11701 (2015)}.

\bibitem {SL} 	S. Liu, L. Z. Mu, and H. Fan,  
\href {https://doi.org/10.1103/PhysRevA.91.042133   }
{        Phys. Rev. A {\bf 91}, 042133 (2015)}.

\bibitem {YLX} Y. L. Xiao, N. H. Jing, S. M. Fei, T. Li, X. Q. Li-Jost, T. Ma, and Z. X. Wang, \href {https://doi.org/10.1103/PhysRevA.93.042125 } {Phys. Rev. A 93, 042125 (2016)}.

\bibitem {FA}	F. Adabi, S. Salimi, and S. Haseli,  
 \href {https://doi.org/10.1103/PhysRevA.93.062123   }
{        Phys. Rev. A {\bf 93}, 062123 (2016)}.


\bibitem {mf11} D. Wang, F. Ming, X. K. Song, L. Ye, and J. L. Chen,
\href {https://doi.org/10.1140/epjc/s10052-020-8403-y    }
{         Eur. Phys. J. C {\bf 80}, 800 (2020)}.

\bibitem {mf} F. Ming, D. Wang, X. G. Fan, W. N. Shi, L. Ye, and J. L. Chen,
\href {https://doi.org/10.1103/PhysRevA.102.012206}
{ Phys. Rev. A {\bf 102}, 012206 (2020)}.

\bibitem {xbf} B. F. Xie, F. Ming, D. Wang, L. Ye, and J. L. Chen,
\href {https://doi.org/10.1103/PhysRevA.104.062204    }
{  Phys. Rev. A {\bf 104}, 062204 (2021)}.

\bibitem {xbf1} L. J. Li, F. Ming, X. K. Song, L. Ye, and D. Wang,
\href {https://doi.org/10.1140/epjc/s10052-022-10687-1} {Eur. Phys. J. C {\bf82}, 726 (2022)}.

\bibitem {LLD} L. Wu, L. Ye, and D. Wang,
\href {https://doi.org/10.1103/PhysRevA.106.062219}
{ Phys. Rev. A {\bf 106}, 062219 (2022)}.

\bibitem {QHZ} Q. H. Zhang and S. M. Fei,
\href {https://doi.org/10.1103/PhysRevA.108.012211} {Phys. Rev. A {\bf 108}, 012211 (2023)}.

\bibitem {TianYu} T. Y. Wang and D. Wang, \href {https://doi.org/10.1016/j.physleta.2024.129364} {Phys. Lett. A {\bf 499}, 129364 (2024)}.

\bibitem {Rp} R. Prevedel, D. R. Hamel, R. Colbeck, K. Fisher, and K. J. Resch, 
\href {https://doi.org/10.1038/nphys2048     }
{   Nat. Phys. {\bf 7}, 757 (2011)}.

\bibitem {Cfl} C. F. Li, J. S. Xu, X. Y. Xu, K. Li, and G. C. Guo, 
\href {https://doi.org/10.1038/nphys2047      }
{   Nat. Phys. {\bf 7}, 752 (2011)}.

\bibitem{ZY} Z. Y. Xu, S. Q. Zhu, W. L. Yang,
\href {https://doi.org/10.1063/1.4771988}{Appl. Phys. Lett.  {\bf101}, 244105 (2012).}

\bibitem {wcm11} W. C. Ma, Z. H. Ma, H. Y. Wang, Z. H. Chen, Y.  Liu, F. Kong, Z. K. Li, X. H. Peng, M. J. Shi, F. Z. Shi, S. M. Fei, and J. F. Du,
\href {https://doi.org/10.1103/PhysRevLett.116.160405         }
{   Phys. Rev. Lett. {\bf 116}, 160405 (2016)}.

\bibitem {zxc}	Z. X. Chen, J. L. Li, Q. C. Song, H. Wang, S. M. Zangi, and C. F. Qiao,  
\href {https://doi.org/10.1103/PhysRevA.96.062123  }
{    Phys. Rev. A {\bf 96}, 062123 (2017)}.

\bibitem {wml} W. M. Lv, C. Zhang, X. M. Hu, H. Cao, J. Wang, and Y. F. Huang, 
\href {https://doi.org/10.1103/PhysRevA.98.062337  }
{    Phys. Rev. A {\bf 98}, 062337 (2018)}.

\bibitem {hyw11}  H. Y. Wang, Z. H. Ma, S. J. Wu, W. Q. Zheng, Z. Cao, Z. H. Chen, Z. K. Li, S. M. Fei, X. H. Peng, V. Vedral, and J. F. Du,
\href {https://doi.org/10.1038/s41534-019-0153-z  }
{     
npj Quantum Information {\bf 5}, 39 (2019)}.

\bibitem {zxc11}  Z. X. Chen, H. Wang, J. L. Li, Q. C. Song, and C. F. Qiao,
\href {https://doi.org/10.1038/s41598-019-42089-x  }
{     
 Sci. Rep. {\bf 9}, 5687 (2019)}.

\bibitem {wcxy1}   W. M. Lv, C. Zhang, X. M. Hu, Y. F. Huang, H. Cao, J. Wang, Z. B. Hou, B. H. Liu, C. F. Li, and G. C. Guo,
\href {https://doi.org/10.1038/s41598-019-45205-z   }
{
    Sci. Rep. {\bf 9}, 8748 (2019)}.

\bibitem {Yong} Z. Y. Ding, H. Yang, D. Wang, H. Yuan, J. Yang, and L. Ye,
\href {https://doi.org/10.1103/PhysRevA.101.032101   }
{   
 Phys. Rev. A {\bf 101}, 032101 (2020)}.



\bibitem {JF} J. Feng, Y. Z. Zhang, M. D. Gould, H. Fan,
 \href {https://doi.org/10.1016/j.physletb.2015.02.058  }
{ Phys. Lett. B {\bf 743}, 198 (2015)}.

\bibitem {JL} J. L. Huang, F. W. Shu, Y. L. Xiao, M. H. Yung,
 \href {https://doi.org/10.1140/epjc/s10052-018-6026-3 }
{Eur. Phys. J. C {\bf 78}, 545 (2018)}.

\bibitem {DWW} D. Wang, W. N. Shi, R. D. Hoehn, F. Ming, W. Y. Sun, S. Kais, L. Ye,
 \href {https://doi.org/10.1002/andp.201800080 }
{Ann. Phys. (Berlin) {\bf 530}, 1800080 (2018)}.

\bibitem {FMDW} F. Ming, D. Wang, L. Ye,
 \href {https://doi.org/10.1002/andp.201900014 }
{Ann. Phys. (Berlin) {\bf 531}, 1900014 (2019)}.

{\bibitem {GWG} G. W. Gibbons,
\href {https://doi.org/10.1016/0550-3213(82)90170-5 }
{ Nucl. Phys. B {\bf 207}, 337 (1982).}}

{\bibitem {GKM} G. W. Gibbons, K. -i. Maeda, \href {https://doi.org/10.1016/0550-3213(88)90006-5 }
{Nucl. Phys. B {\bf 298}, 741 (1988).}}






{\bibitem {RKRB} {R. Kerner, R.B. Mann,}
\href {https://doi.org/10.1103/PhysRevD.73.104010 }
{Phys. Rev. D {\bf 73},  104010 (2006).}}

\end{references}
\end{document}